\documentclass[runningheads]{llncs}
\usepackage[T1]{fontenc}
\usepackage{graphicx}
\usepackage[misc,geometry]{ifsym}
\usepackage{amsmath}
\usepackage{color}
\usepackage{multirow}
\begin{document}
\title{Fine-Grained Unsupervised Cross-Modality Domain Adaptation for Vestibular Schwannoma Segmentation}

\titlerunning{Fine-Grained Unsupervised Cross-Modality Domain Adaptation}
%
\author{Luyi Han\inst{1,2} \and
Tao Tan\inst{2,3}\textsuperscript{(\Letter)} \and
Ritse Mann\inst{1,2} 
}
\authorrunning{L. Han et al.}
%
\institute{Department of Radiology and Nuclear Medicine, Radboud University Medical Center, Geert Grooteplein 10, 6525 GA, Nijmegen, The Netherlands.\\
\and
Department of Radiology, The Netherlands Cancer Institute, Plesmanlaan 121, 1066 CX, Amsterdam, The Netherlands.\\
\and
Faculty of Applied Science, Macao Polytechnic University, 999078, Macao, China\\
\email{\{taotanjs\}@gmail.com}
}
\maketitle              
\begin{abstract}
The domain adaptation approach has gained significant acceptance in transferring styles across various vendors and centers, along with filling the gaps in modalities. However, multi-center application faces the challenge of the difficulty of domain adaptation due to their intra-domain differences. We focus on introducing a fine-grained unsupervised framework for domain adaptation to facilitate cross-modality segmentation of vestibular schwannoma (VS) and cochlea.
We propose to use a vector to control the generator to synthesize a fake image with given features. And then, we can apply various augmentations to the dataset by searching the feature dictionary. The diversity augmentation can increase the performance and robustness of the segmentation model.
On the CrossMoDA validation phase Leaderboard, our method received a mean Dice score of 0.765 and 0.836 on VS and cochlea, respectively.

\end{abstract}
\section{Introduction}
The use of domain adaptation in clinical settings has become increasingly popular to enhance the effectiveness of deep learning methods.
The objective of the Cross-Modality Domain Adaptation (CrossMoDA) challenge~\cite{dorent2023crossmoda} is to accurately segment two brain structures: the vestibular schwannoma (VS) and cochlea.
Although contrast-enhanced T1 (ceT1) MR imaging is a common diagnostic and surveillance tool for patients with VS, the research on non-contrast imaging, such as T2-weighted imaging (T2), is increasing due to its lower risk and cost-effectiveness.
To prevent the need for extra annotation, CrossMoDA works towards transferring acquired knowledge from ceT1 images to T2 images through the creation of domain adaptation between unpaired ceT1 and T2.
The use of multi-center data from London SC-GK (ldn), Tilburg SC-GK (etz), and UK MC-RC (ukm) poses a challenge in domain adaptation due to intra-domain bias. Simply transferring images from the ceT1 domain to the T2 domain using a single model results in a loss of diversity in the multi-center T2 images. Thus, we propose a fine-grained domain adaptation model controlled with more specific features to synthesize T2 images.

\section{Related Work}
Extensive validation of unsupervised domain adaptation for the vestibular system and cochlea segmentation has been conducted through the CrossMoDA challenge~\cite{dorent2023crossmoda}. In present studies, an image-to-image translation technique, such as CycleGAN~\cite{zhu2017unpaired}, is utilized to produce pseudo-target domain images from source domain images. These generated images and their corresponding manual annotations are subsequently employed to train the segmentation models.
Several recent studies have utilized varying versions of CycleGAN to convert ceT1 images into hrT2 images for brain tumor segmentation. The main objective of these studies is to maintain the structural integrity of the low-intensity regions in real hrT2 scans, such as the VS and cochleas. Specifically, Dong et al.~\cite{dong2021unsupervised} utilize NiceGAN~\cite{chen2020reusing}, which employs discriminators for encoding, while Choi et al.~\cite{choi2021using} use CUT~\cite{park2020contrastive} and post-processing techniques. Furthermore, Shin et al.~\cite{shin2022cosmos} implement an additional decoder and an iterable self-training strategy.
Kang et al.~\cite{kang2023multi} utilize two image translation models in parallel, each incorporating a pixel-level consistent constraint and a patch-level contrastive constraint, respectively.
Sallé et al.~\cite{salle2023cross} propose tumor blending augmentation combined with CycleGAN, which diversifies target regions of interest during training, improving the segmentation generalization of the model during testing.
Han et al.~\cite{han2022unsupervised} acquire a shared representation from both ceT1 and hrT2 images alongside the successful recovery of another modality from the latent representation. In addition, they utilized proxy tasks of VS segmentation and brain parcellation to regulate the consistency of image structures in domain adaptation effectively.

\begin{figure}
    \centering
    \includegraphics[width=1\linewidth]{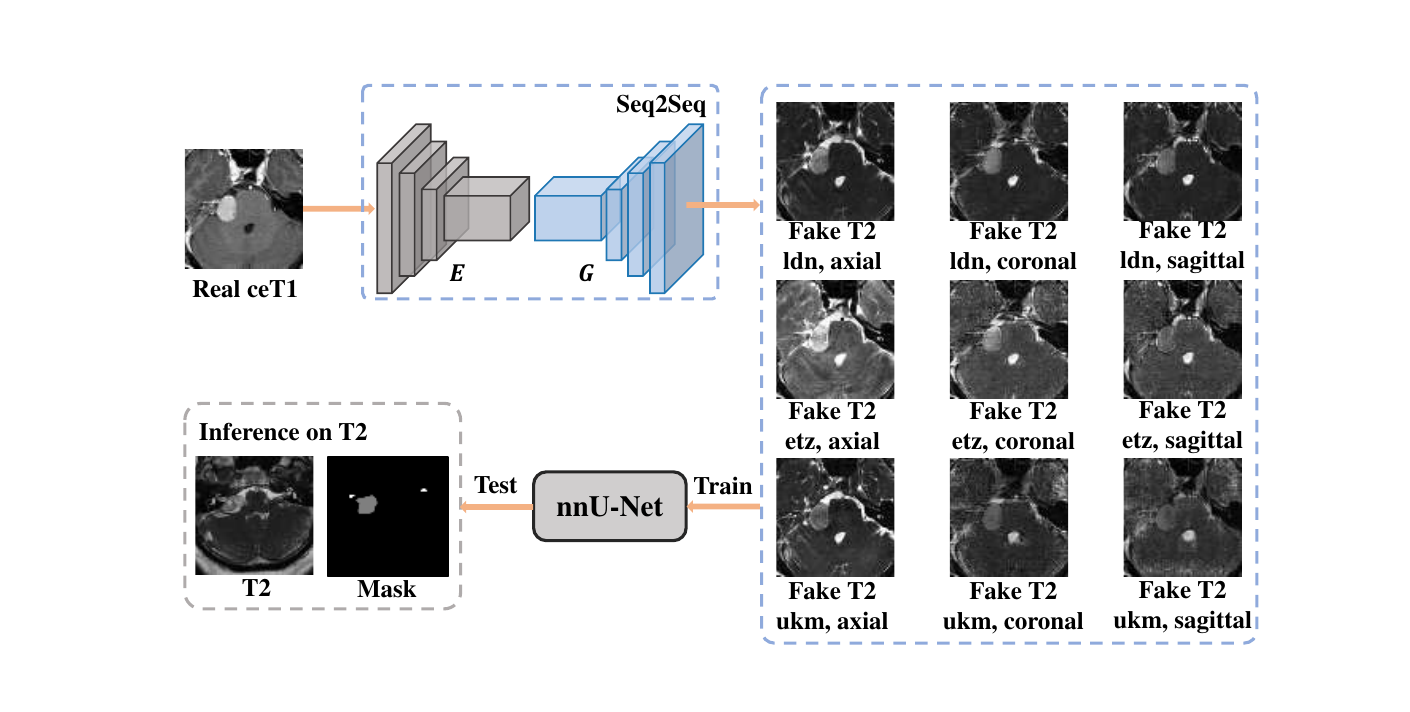}
    \caption{Overview of the proposed fine-grained unsupervised domain adaptation segmentation framework. The plane (axial, coronal, and sagital) presents the plane of 2D slices for the input image to the Seq2Seq model. The plane (axial, coronal, and sagittal) presents the plane of 2D slices for the image to input to the Seq2Seq model. The SeqSeq model can only process 2D slices and re-stack these slices into 3D images. For better comparison, we only show the axial plane of these images.}
    \label{fig:framework}
\end{figure}

\section{Method}
\subsection{Framework Overview}
Figure~\ref{fig:framework} illustrates the proposed fine-grained unsupervised domain adaptation segmentation framework. We first employ a conditional generator (Seq2Seq)~\cite{han2023synthesis} to synthesize the corresponding T2 image from a given ceT1 image. The T2 image can be augmented by changing the feature codes used to control the Seq2Seq generator.
Using the fake T2 dataset after augmentation, we can train a robust segmentation network -- the nnU-Net~\cite{isensee2021nnu} model.

\begin{figure}
    \centering
    \includegraphics[width=1\linewidth]{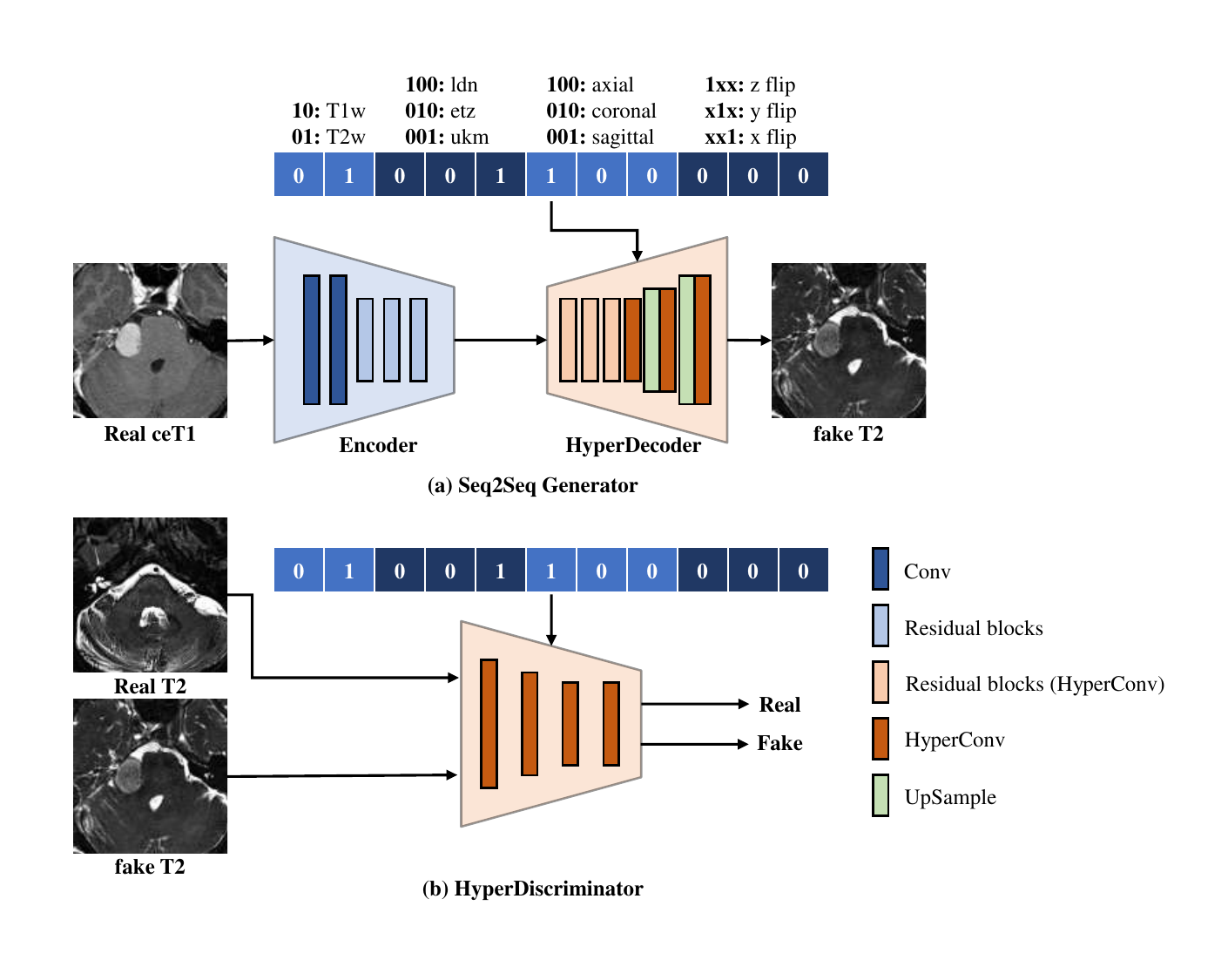}
    \caption{The architecture of (a) Seq2Seq generator and (b) HyperDiscriminator. The given conditional zero-one code can control the synthesized image of the generator. The conditional code is the combination of target modality (T1w or T2w), center (ldn, etz, ukm), plane (axial, coronal, sagittal), and flip augmentation (z-axis, y-axis, x-axis).}
    \label{fig:architecture}
\end{figure}

\subsection{Fine-grained domain adaptation}
Figure~\ref{fig:architecture} illustrates the architecture of the Seq2Seq generator~\cite{han2023synthesis} and HyperDiscriminator. The seq2Seq generator is a conditional autoencoder-like model, inputting with a real ceT1 image and a conditional zero-one code and outputting with the fake T2 image. The conditional code combines the target modality, center, plane, and flip augmentation code. Note that, the plane in the conditional code presents the plane of 2D slices for the input image because the proposed model can only handle 2D slices. Furthermore, we use the HyperDiscriminator to discriminate whether the image is real or fake. We train our models by following the training process and loss function in \cite{han2022unsupervised}, and keep all the hyper-parameters the same.

\subsection{Segmentation with domain augmentation}
With the proposed Seq2Seq, we can control the intra-domain diversity of the fake T2 images. Specifically, we augment each real ceT1 image into nine fake images with the combination of the image center and plane view as shown in Fig.~\ref{fig:framework}. Finally, we train the nnU-Net~\cite{isensee2021nnu} model feeding with the augmented dataset.

\section{Experimental Results} \label{section:experiment}
\subsection{Materials and Implementation Details}
\subsubsection{Dataset.}
The dataset for the CrossMoDA challenge is an extension of the publicly available Vestibular-Schwannoma-SEG collection~\cite{wijethilake2022boundary,kujawa2022deep}, which is divided into the training dataset (227 subjects with ceT1 images and other unpaired 295 subjects with T2 images) and the validation dataset (96 subjects with T2 images).
All imaging datasets were manually segmented for cochlea and VS, and the VS segmentation was split into the intra- and extra-meatal regions.

\subsubsection{Data preprocessing.}
The pipeline of image preprocessing is modified from MSF-Net~\cite{han2022unsupervised}. All the images are first resampled to the spacing of $0.4102\times0.4102\times0.4102$ mm due to various spacing of images from different centers. Then we normalize images to $[0,1]$ by setting the intensity range from 0 to 99.5 percentile. We select an identical ceT1 image as the atlas to improve domain adaptation performance and employ intra- and inter-modality affine transformation on all the ceT1 and T2 images, respectively. Here, we utilize mutual information (MI) loss for T2 images and normalized cross-correlation (NCC) loss for ceT1 images. Finally, based on the distribution of tumor areas in the training set, we crop the images to the size of $256\times256\times256$ by setting a fixed region.
Finally, 2D slices selected from axial, coronal, and sagittal planes are used to train the models.

\subsubsection{Implementation Details.}
We implemented our method using PyTorch with NVIDIA 3090 RTX.
The encoder architecture of Seq2Seq is comprised of three convolutional layers and six residual blocks. The first convolutional layer is responsible for encoding the intensities into features, while the second and third convolutional layers conduct four-time downsampling of images. The high-level representation is then extracted by the residual blocks, whose channels are of sizes 64, 128, 256, and 256, respectively. The HyperDecoder of Seq2Seq is an inverse architecture of the encoder, which has all the convolutional layers replaced with HyperConv. The HyperDiscriminator is implemented with four HyperConv layers with a kernel size of 4, stride of 2, and output channel of 64, 128, 256, and 512, respectively, followed by a convolutional layer with a kernel size of 1 and an output channel of 1.
The Seq2Seq generator is trained with Adam optimizer, a learning rate of $2\times10^{-4}$, a default of 1,000 epochs, and a batch size of 1. nnU-Net is trained with its default 3D full-resolution settings.

\begin{figure*}[!htbp]
    \begin{minipage}{0.1\linewidth}
        \centerline{T2}
    \end{minipage}
    \begin{minipage}{0.14\linewidth}
        \centerline{\includegraphics[width=\textwidth]{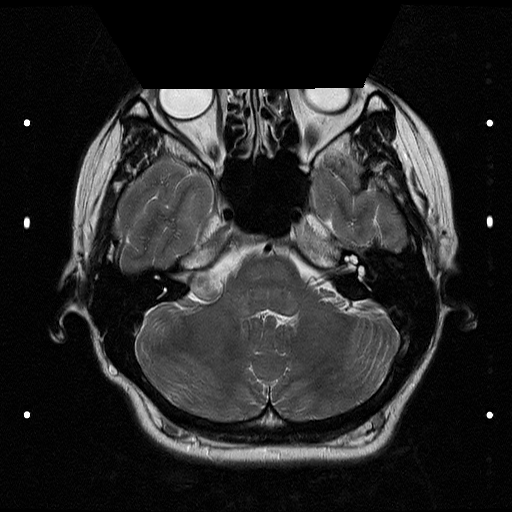}}
    \end{minipage}
    \begin{minipage}{0.14\linewidth}
        \centerline{\includegraphics[width=\textwidth]{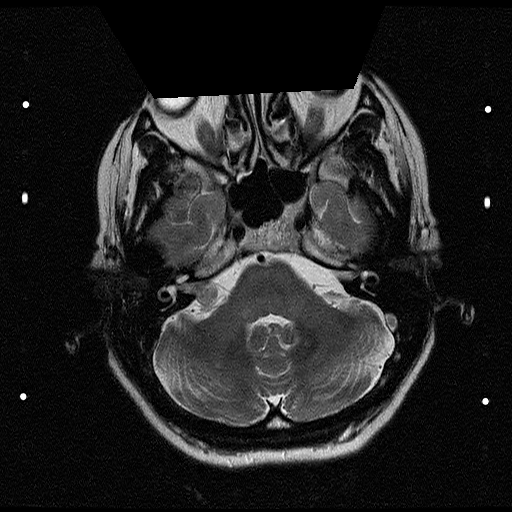}}
    \end{minipage}
    \begin{minipage}{0.14\linewidth}
        \centerline{\includegraphics[width=\textwidth]{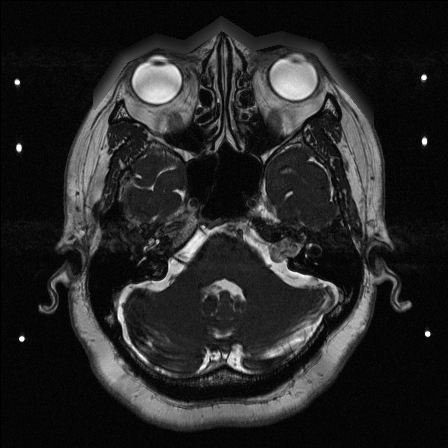}}
    \end{minipage}
    \begin{minipage}{0.14\linewidth}
        \centerline{\includegraphics[width=\textwidth]{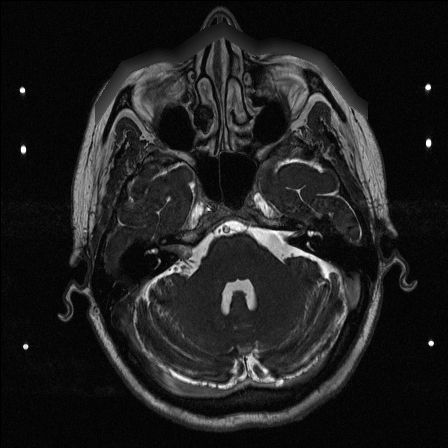}}
    \end{minipage}
    \begin{minipage}{0.14\linewidth}
        \centerline{\includegraphics[width=\textwidth]{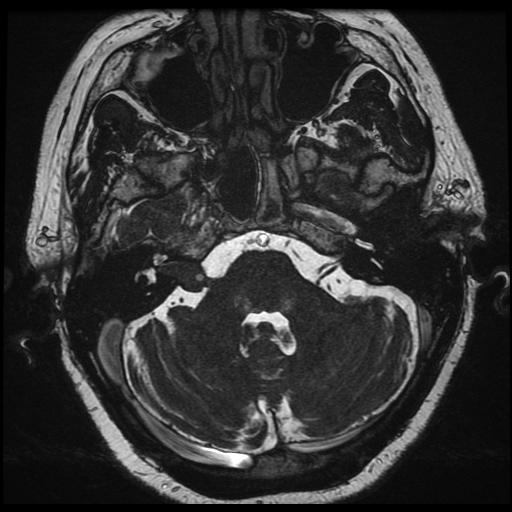}}
    \end{minipage}
    \begin{minipage}{0.14\linewidth}
        \centerline{\includegraphics[width=\textwidth]{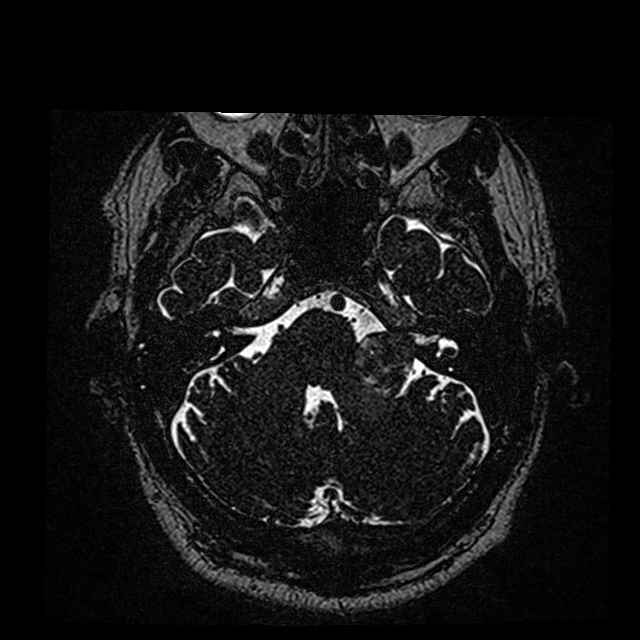}}
    \end{minipage}
    
    \vspace{2pt}
    
    \begin{minipage}{0.1\linewidth}
        \centerline{Mask}
    \end{minipage}
    \begin{minipage}{0.14\linewidth}
        \centerline{\includegraphics[width=\textwidth]{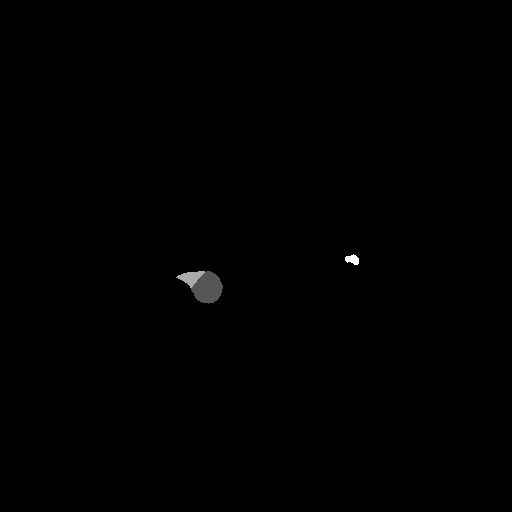}}
    \end{minipage}
    \begin{minipage}{0.14\linewidth}
        \centerline{\includegraphics[width=\textwidth]{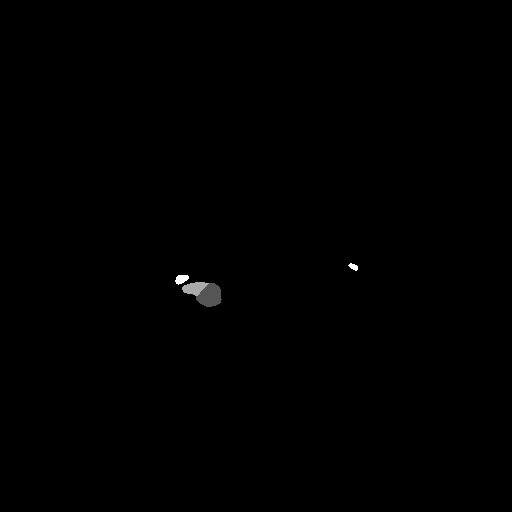}}
    \end{minipage}
    \begin{minipage}{0.14\linewidth}
        \centerline{\includegraphics[width=\textwidth]{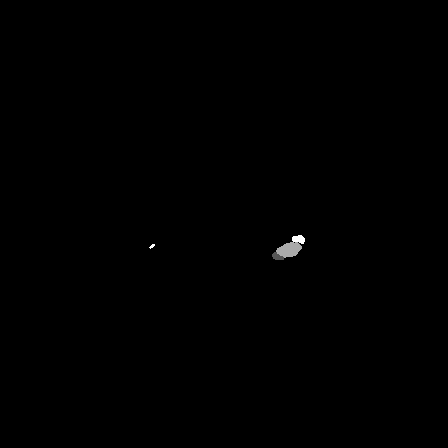}}
    \end{minipage}
    \begin{minipage}{0.14\linewidth}
        \centerline{\includegraphics[width=\textwidth]{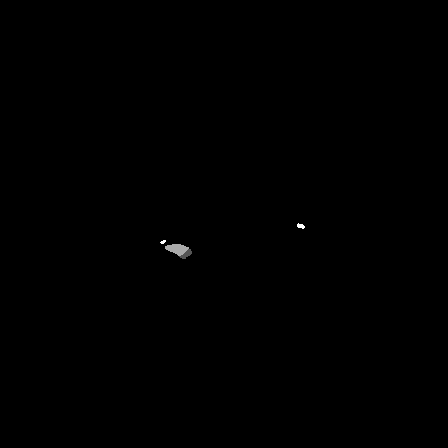}}
    \end{minipage}
    \begin{minipage}{0.14\linewidth}
        \centerline{\includegraphics[width=\textwidth]{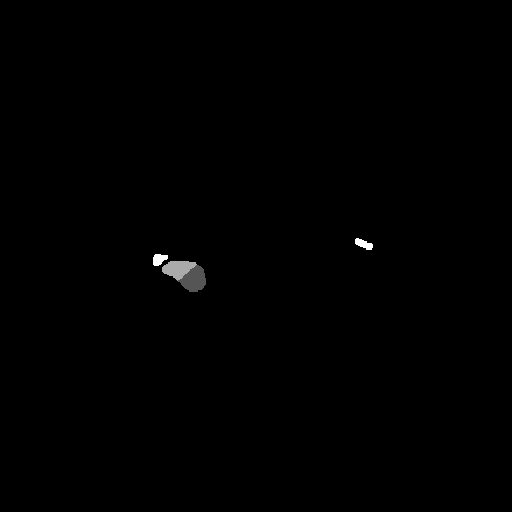}}
    \end{minipage}
    \begin{minipage}{0.14\linewidth}
        \centerline{\includegraphics[width=\textwidth]{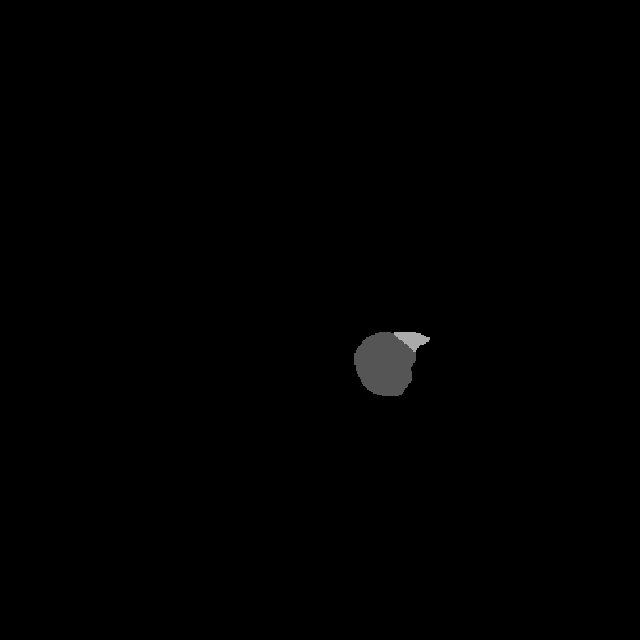}}
    \end{minipage}

    \begin{minipage}{0.1\linewidth}
        \centerline{}
    \end{minipage}
    \begin{minipage}{0.14\linewidth}
        \centerline{Case 1}
    \end{minipage}
    \begin{minipage}{0.14\linewidth}
        \centerline{Case 2}
    \end{minipage}
    \begin{minipage}{0.14\linewidth}
        \centerline{Case 3}
    \end{minipage}
    \begin{minipage}{0.14\linewidth}
        \centerline{Case 4}
    \end{minipage}
    \begin{minipage}{0.14\linewidth}
        \centerline{Case 5}
    \end{minipage}
    \begin{minipage}{0.14\linewidth}
        \centerline{Case 6}
    \end{minipage}
	
    \caption{Examples of vestibular schwannoma (VS) and cochlea segmentation.} \label{fig:domainadaptation}
\end{figure*}

\subsection{Results}
Fig.~\ref{fig:domainadaptation} shows the performance of VS and cochlea segmentation. With fine-grained modality augmentation, the segmentation model can correctly segment images from different centers.
We compare our proposed method with MSF-Net~\cite{han2022unsupervised} on ceT1 and T2 domain adaptation to evaluate the influence of different generation performances for further segmentation.
Table~\ref{tab:seg} and Table~\ref{tab:seg2} show the segmentation results for the nnU-Net models training with fake T2 images generated by different methods and whether augmented with varying styles of planes and centers. The proposed method achieves better results than MSF-Net and Seq2Seq on the validation set, and the ablation study shows that more augmentation can increase the generalization ability of the segmentation model.

\begin{table}
\centering
\caption{Segmentation results of VS and cochlea for nnU-Net utilizing generated T2 images with different domain adaptation methods. The best result is in bold.}\label{tab:seg}
\begin{tabular}{lccccc}
\hline
\multirow{2}*{Methods}
    & \multicolumn{2}{c}{VS} && \multicolumn{2}{c}{Cochlea} \\
    \cline{2-3}\cline{5-6}
    & Dice $\uparrow$ & ASSD $\downarrow$ && Dice $\uparrow$ & ASSD $\downarrow$ \\
\hline
MSF-Net~\cite{han2022unsupervised} & 0.671$\pm$0.304 & 10.6$\pm$38.4 && 0.820$\pm$0.039 & 0.300$\pm$0.371 \\
Seq2Seq~\cite{han2023synthesis} & 0.702$\pm$0.282 & 9.29$\pm$38.3 && 0.804$\pm$0.040 & 0.275$\pm$0.136 \\
Seq2Seq+Plane & 0.710$\pm$0.275 & 8.74$\pm$37.9 && 0.833$\pm$0.034 & 0.229$\pm$0.128 \\
Seq2Seq+Plane+Center & \textbf{0.765$\pm$0.255} & \textbf{7.49$\pm$37.4} && \textbf{0.836$\pm$0.031} & \textbf{0.218$\pm$0.127} \\
\hline
\end{tabular}
\end{table}

\begin{table}
\centering
\caption{Segmentation results of intra- and extra-meatal region of VS for nnU-Net utilizing generated T2 images with different domain adaptation methods. The best result is in bold.}\label{tab:seg2}
\begin{tabular}{lccccc}
\hline
\multirow{2}*{Methods}
    & \multicolumn{2}{c}{Intra-Meatal} && \multicolumn{2}{c}{Extra-Meatal} \\
    \cline{2-3}\cline{5-6}
    & Dice $\uparrow$ & ASSD $\downarrow$ && Dice $\uparrow$ & ASSD $\downarrow$ \\
\hline
MSF-Net~\cite{han2022unsupervised} & 0.542$\pm$0.277 & 11.1$\pm$38.7 && 0.706$\pm$0.273 & 14.6$\pm$64.0 \\
Seq2Seq~\cite{han2023synthesis} & 0.561$\pm$0.264 & 9.77$\pm$38.6 && 0.750$\pm$0.222 & \textbf{5.98$\pm$37.7} \\
Seq2Seq+Plane & 0.550$\pm$0.267 & 16.1$\pm$61.7 && 0.757$\pm$0.227 & 12.7$\pm$63.7 \\
Seq2Seq+Plane+Center & \textbf{0.598$\pm$0.242} & \textbf{7.67$\pm$37.4} && \textbf{0.813$\pm$0.174} & 8.56$\pm$52.4 \\
\hline
\end{tabular}
\end{table}

\section{Discussion}
According to the findings depicted in Fig.~\ref{fig:domainadaptation}, despite noticeable differences in the visual appearance of T2 images from different centers due to variations in scanning devices and parameters, our segmentation model has accomplished the task of accurately segmenting VS and cochlea. This outcome highlights the efficacy of data augmentation through fine-granted domain adaptation in enhancing the generalization of subsequent segmentation models. In addition, as indicated by the results presented in Table~\ref{tab:seg} and Table~\ref{tab:seg2}, generating images from diverse planes of view and multiple centers through augmentation can enhance the segmentation precision of VS and cochlear.
Nevertheless, due to the model being able only to process 2D slices and the absence of spatial 3D information extraction, the resulting fake T2 images inadequately display the morphology of certain minute tumors, ultimately failing to segment these smaller tumors accurately.

\section{Conclusion}
In this study, we propose a method for fine-grained unsupervised cross-modality domain adaptation. Our approach enables the synthesis of T2 images from different centers, thereby increasing the generalization of the segmentation model. Our proposed method achieves a third rank during the test phase for cross-modal VS (intra- and extra-meatal region) and cochlear segmentation in the CrossMoDA 2023 challenge.

\subsubsection{Acknowledgement}
Luyi Han was funded by Chinese Scholarship Council (CSC) scholarship.

%
\bibliographystyle{splncs04}
\bibliography{refs}
	
\end{document}